\documentclass[12pt]{amsart}
\usepackage{graphicx}
\usepackage[centertags]{amsmath}
\usepackage{amsfonts}
\usepackage{amssymb}
\usepackage{amsthm}

\newtheorem{thm}{Theorem}%[section]
\newtheorem{lem}{Lemma}[section]
\newtheorem{cor}[lem]{Corollary}

\newtheorem{prop}[lem]{Proposition}
\theoremstyle{definition}

\theoremstyle{remark}
\newtheorem{rem}{Remark}[section]
\numberwithin{equation}{section}

\newcommand{\norm}[1]{\left\Vert#1\right\Vert}

\newcommand{\set}[1]{\left\{#1\right\}}

\newcommand{\To}{\longrightarrow}

\newcommand{\calA}{\mathcal{A}}

\newcommand{\calU}{\mathcal{U}}

\newcommand{\calW}{\mathcal{W}}

\newcommand{\bbZ}{\mathbb{Z}}

\newcommand{\bbQ}{\mathbb{Q}}
\newcommand{\bbR}{\mathbb R}

\newcommand{\bbN}{\mathbb N}
\newcommand{\bbT}{\mathbb{T}}
\newcommand{\Tr}{ \mbox{Tr}}
\newcommand{\Op}{ \operatorname{Op}}

\newcommand{\GL}{ \mbox{GL}}

\newcommand{\Sp}{\mathrm{Sp}}

\newcommand{\rank}{\mathrm{rank }}
\newcommand{\supp}{\mathrm{supp}}

\newcommand{\calH}{\mathcal{H}}

\begin{document}
\title[Scarring on invariant manifolds]
{Scarring on invariant manifolds for perturbed quantized hyperbolic toral automorphisms}%
\author{Dubi Kelmer }%
\address{Raymond and Beverly Sackler School of Mathematical Sciences,
Tel Aviv University, Tel Aviv 69978, Israel}
\email{kelmerdu@post.tau.ac.il}

\thanks{}%

\date{\today}%
\dedicatory{}%
\commby{}%

\begin{abstract}
We exhibit scarring for certain nonlinear ergodic toral
automorphisms. There are perturbed quantized hyperbolic toral
automorphisms preserving certain co-isotropic submanifolds. The
classical dynamics is ergodic, hence in the semiclassical limit
almost all eigenstates converge to the volume measure of the
torus. Nevertheless, we show that for each of the invariant
submanifolds, there are also eigenstates which localize and
converge to the volume measure of the corresponding submanifold.
\end{abstract}

\maketitle
\section{Introduction}
    A significant problem in quantum chaos, is to understand the behavior of eigenstates of classically chaotic
    systems in the semiclassical limit. In particular, one would like
    to classify the possible measures on phase space obtained as a
    quantum limit (e.g., in terms of the corresponding Wigner distributions).

    The main result in this direction is the
    \v{S}hnirel'man theorem (also referred to as the
    Quantum Ergodicity Theorem). This theorem states that for
    classically ergodic systems almost all sequences of
    eigenstates converge to the volume measure on the corresponding energy shell  ~\cite{BD,DD,S,Z}.
    Going beyond the \v{S}hnirel'man theorem, one would like to
    classify what other possible invariant measures (if any) can be
    obtained as a quantum limit.

    For surfaces of constant negative
    curvature, the Quantum Unique Ergodicity conjecture suggests that
    the only possible limiting measure is the volume measure \cite{RS}. This
    conjecture has been proved for the case of compact arithmetic
    surfaces, if one takes into account the arithmetic symmetries of the system \cite{EL}.

    A similar situation also occurs for quantized linear symplectic maps of the two dimensional torus.
    Here again, the quantized system exhibits arithmetic
    symmetries. After taking these symmetries into account, the
    only possible limiting measure is shown to be the volume
    measure \cite{KR}. On the other hand, when considering eigenstates of
    the propagator without the symmetries, one can construct a
    thin sequence of eigenstates partially localized on periodic orbits
    \cite{FND}.

    One can consider also quantized linear symplectic maps of higher
    dimensional tori. For such a system, if the classical map leaves no invariant rational isotropic subspaces,
    the only limiting measure
    (after taking the arithmetic symmetries into account) is again the volume measure on the whole torus \cite{K}.
    However, if there are rational isotropic invariant subspaces, then there are sequences of quantum states
    localized on corresponding co-isotropic invariant manifolds.
    Furthermore, this phenomenon is stable under the arithmetic symmetries of the system \cite{K}.
    In these notes we show that this localization is also stable
    under certain nonlinear perturbations, which preserve these
    co-isotropic manifolds.

    We briefly review the basic setup:
    Consider a discrete time dynamical system given by
    iterating the action of a symplectic linear map $A\in\Sp(2d,\bbZ)$ on
    $\bbT^{2d}=\bbR^{2d}/\bbZ^{2d}$. Assume that $A$ has no eigenvalues of modulus $1$,
    so that the map is Anosov and hence stably ergodic.
    Further assume, that the action of $A$ on $\bbQ^{2d}$ leaves an invariant isotropic
    subspace $V\subseteq \bbQ^{2d}$ of dimension $1<d_0\leq d$. Denote by $\Lambda=V\cap\bbZ^{2d}$
    the integral points of $V$, then for every point $\xi\in\bbT^{2d}$ which is fixed by $A$ the manifold
    \[X_\xi=\set{x\in\bbT^{2d}|e_n(x)=e_n(\xi),\;\forall
    n\in\Lambda},\]
    is a closed co-isotropic submanifold that is invariant under the action of $A$.
    Let $\phi_H$ be a Hamiltonian map on $\bbT^{2d}$ which leaves all the manifolds $X_\xi$
    invariant, and look at a perturbation $\Phi=A\circ \phi_H$  (also leaving the $X_\xi$'s invariant).
    Furthermore, we can make the
    perturbation sufficiently small so that the perturbed map remains ergodic.

    For quantum mechanics on $\bbT^{2d}$, the admissible values of
    Planck's constant are inverses of integers $h=1/N$, and the space of
    quantum states is then $\calH_N=L^2[(\bbZ/N\bbZ)^d]$.
    The semiclassical limit is achieved by taking $N\rightarrow
    \infty$. For $f\in C^\infty(\bbT^{2d})$  a smooth observable, we denote by
    $\Op_N(f):\calH_N\rightarrow \calH_N$ its quantization. Any
    quantum state $\psi\in\calH_N$ can then be interpreted as a
    distribution on $\bbT^{2d}$ via the Wigner distribution
    $\calW_N(\psi)$ sending any smooth
    $f$ to its expectation value $\langle \Op_N(f)\psi,\psi\rangle$.
    The quantization of the map $\Phi$, is a family of unitary operators $\calU_N(\Phi)$
    acting on $\calH_N$, satisfying the Egorov identity in the
    semiclassical limit.
    %\[\norm{\calU_N(\Phi)^{-1}\Op_N(f)\calU_N(\Phi)-\Op_N(f\circ
    %\phi)}=O_{f}(\frac{1}{N})\]

    A measure $\mu$ on $\bbT^{2d}$ is called a
    limiting quantum measure, if there is a sequence
    $\psi=\psi^{(N)}\in\calH_N$ of eigenstates of $\calU_N(\Phi)$, such
    that as $N\to\infty$, the corresponding Wigner distributions
    $\calW_N(\psi)$ converge (weak$^*$) to $\mu$.
    We can now state the main theorems,
    establishing that the volume measures of the manifolds $X_\xi$ are all limiting quantum measures.

    For each fixed point $\xi$ and any integer $N$ divisible by
    $R=\det(A-I)$, we define
    \[\calH_{N,\xi}=\set{\psi\in\calH_N| \Op_N(e_n)\psi=e_n(\xi)\psi,\;\forall n\in\Lambda}.\]
    \begin{thm}\label{tSCARR}
    The spaces $\calH_{N,\xi}$ are of dimension $N^{d-d_0}$ and
    are invariant under $\calU_N(\Phi)$.
    Furthermore, if $\psi_{j}\in \calH_{N_j,\xi}$ is a sequence of states such that
    the Wigner distributions
    converge
    to some measure $\mu$ on $\bbT^{2d}$,
    then $\mu$ is supported on $X_{\xi}$.
    \end{thm}

    The spaces $\calH_{N,\xi}$ are invariant under
    $\calU_N(\Phi)$, and hence have a basis composed of eigenstates.
    We want to show that there is a sequence of such
    eigenstates, converging to the volume measure of $X_\xi$.
    In fact, we show that there are many such sequences.
    \begin{thm}\label{tERG}
    For each $N$ (divisible by $R$) take an orthonormal basis $\psi_i=\psi_i^{(N)}$ of
    $\calH_{N,\xi}$ composed of eigenstates of $\calU_N(\Phi)$.
    Then there are subsets $S_N\subset\{1,\ldots, N^{d-d_0}\}$ satisfying $\lim_{N\to\infty}\frac{|S_N|}{N^{d-d_0}}=1$,
    such that for any sequence $\{\psi_{i_N}\}$ with $i_N\in S_N$
    the corresponding Wigner distributions
    $\calW_N(\psi_{i_N})$,
    converge to the volume measure $dm_{X_\xi}$ concentrated on $X_\xi$.
    \end{thm}

    \begin{rem}
    In the case that the fixed point $\xi=0$, the
    requirement that $N$ is divisible by $R$ is not necessary, and theorems
    \ref{tSCARR},\ref{tERG} hold for any sequence of $N\to \infty$.
    \end{rem}

\section*{acknowledgments}
    I thank Jens Marklof for suggesting to
    extend the localization on invariant manifolds for perturbed
    maps. This work was supported in part by the Israel Science Foundation founded by
    the Israel Academy of Sciences and Humanities. This work was
    carried out as part of the author's Ph.D. thesis at Tel Aviv
    University, under the supervision of Prof. Zeev Rudnick.

\section{Background: Quantum maps on the torus}\label{sQUANT}
    The full details for the quantization of the cat map and its perturbations on $\bbT^{2}$ can
    be found in ~\cite{BD,DEG,HB,Knabe}. The
    generalization of these procedures for higher dimensions are analogous and are accounted for in
    \cite{BD1,BD3,K,RSO}. We give a short review of these quantization procedures.
\subsection{Classical dynamics}\label{sCD}
    We consider a discrete time dynamical system given by the
    iteration of a symplectic map $\Phi:\bbT^{2d}\to\bbT^{2d}$.
    More precisely, we consider a map $\Phi=A\circ \phi_H$ that is
    a composition of a linear symplectic map $A\in\Sp(2d,\bbZ)$
    acting on $\bbT^{2d}$ and a Hamiltonian map $\phi_H$ (i.e,
    the evaluation at time one of some Hamiltonian flow).

    The action of $A\in\Sp(2d,\bbZ)$ on the torus
    $\bbT^{2d}=\bbR^{2d}/\bbZ^{2d}$ is induced by
    the natural left action of $A$ on the linear space
    $\bbR^{2d}$, that is $x=(\begin{array}{c }p\\ q\end{array})\in\bbT^{2d}\mapsto Ax$
    (this is well defined since $A$ preserves $\bbZ^{2d}$).
    The induced action is invertible and area preserving.
    Furthermore, if $A$ has no eigenvalues that are roots of unity
    then the induced dynamics are ergodic and mixing.
    If in addition there are no eigenvalues of modulus $1$, then
    the dynamics is of Anosov type and in particular it is stably ergodic
    (c.f. \cite{An,YP}).

    For $H\in C^\infty(\bbT^{2d})$ a smooth real valued function on $\bbT^{2d}$,
    the Hamiltonian flow $\phi_H^t:\bbT^{2d}\to\bbT^{2d}$
    satisfies the differential equations
    \[\frac{d}{dt}(f\circ \phi_H^t)=\{H,f\}\circ \phi_H^t=\{H,f\circ \phi_H^t\},\]
    for any smooth $f\in C^\infty(\bbT^{2d})$, where
    $\{f,g\}=\sum_j
     (\frac{\partial f}{\partial p_j}\frac{\partial g}{\partial q_j}-
     \frac{\partial f}{\partial q_j}\frac{\partial g}{\partial p_j})$ is the Poisson
    brackets.
    The map $\phi_H=\phi_H^1$ is the evaluation at time $t=1$ of this flow.
    Consequently, the dynamics induced from the map $\phi_H$ on
    $\bbT^{2d}$, is just the evaluation of the corresponding Hamiltonian flow at integral times.

    As long as the perturbation is sufficiently small, the perturbed map $\Phi$ remains an Anosov
    diffeomorphism. In particular, after the perturbation the dynamics is again Ergodic and mixing.

\subsection{Quantum mechanics on the torus}\label{sQM}
    For doing quantum mechanics on the torus $\bbT^{2d}$, the admissible values for Plank's
    constant are inverse of integers $h=1/N$. The Hilbert space of states is finite dimensional of dimension
    $N^{d}$. It is convenient to think of it as a space of functions
    $\mathcal{H}_N=L^2((\bbZ/N\bbZ)^d)$, with inner product given
    by:
    \[\langle\varphi,\psi\rangle=\frac{1}{N^d}\sum_{Q\in(\bbZ/N\bbZ)^d}\varphi(Q)\overline{\psi(Q)}.\]

    To any classical observable, $f\in\ C^\infty(\bbT^{2d})$, we assign a corresponding quantum observable
    $\Op_N(f)$ acting on $\calH_N$ by an analog of the Weyl quantization. %a linear operators acting on $\calH_N$.
    We define the operators $\Op_N(f)$ on the
    Fourier basis $e_n(x)=e(n\cdot x),\;n=(n_1,n_2)\in\bbZ^{2d}$ and then expand by linearity.

    For $e_n\in C^\infty(\bbT^{2d})$, the corresponding quantum observables act on $\psi\in \mathcal{H}_N$ via:
    \begin{equation}\label{eTN:1}
    \Op_N(e_n)\psi(Q)=e(\frac{n_1\cdot n_2}{2N})e(\frac{n_2\cdot
    Q}{N})\psi(Q+n_1).
    \end{equation}
    For any smooth classical observable $f\in C^\infty(\mathbb{T}^{2d})$
    with Fourier expansion
    $f(x)=\sum_{n\in\bbZ^{2d}}\hat{f}(n)e_n(x)$ we thus define
    \[\Op_N(f)=\sum_{n\in\bbZ^{2d}}\hat{f}(n)\Op_N(e_n).\]

    The main properties of the elementary observables $\Op_N(e_n)$
    are summarized in the following proposition.
   \begin{prop}\label{pTN:1}
   For any $n,m\in\bbZ^{2d}$
     \begin{enumerate}
     \item The operator $\Op_N(e_n)$ is a unitary operator.
     \item The composition
       \[\Op_N(e_m)\Op_N(e_n)=e(\frac{\omega(m,n)}{2N}) \Op_N(e_{n+m}),\]
       where $\omega(m,n)=m\begin{pmatrix} 0 & I \\ -I & 0\end{pmatrix} n^t$ is the standard symplectic inner product.
     \item The operator $\Op_N(e_n)$ is only dependent on $n\pmod{2N}$.
     \end{enumerate}
   \end{prop}
   These properties are easily derived from the action given in (\ref{eTN:1}).
   Furthermore, they imply that $\Op_N(f)^*=\Op_N(\bar f)$, and in particular for real valued
    functions these operators are Hermitian.

    From the commutation relations of the elementary operators and
    the fast decay of the Fourier coefficients one can obtain that
    as $N\to\infty$,
    \begin{equation}\label{ePROD}
   \norm{\Op_N(f)\Op_N(g)-\Op_N(fg)}=O_{f,g}(\frac{1}{N})
    \end{equation}
   %\[\norm{\Op_N(f)\Op_N(g)-\Op_N(fg)}=O_{f,g}(\frac{1}{N})\]
  %  \[\norm{[\Op_N(f),\Op_N(g)]-i\hbar\Op_N(\{f,g\})}=O_{f,g}(\frac{1}{N})\]

\subsection{Quantum dynamics}
    For $\phi_H:\bbT^{2d}\to\bbT^{2d}$ a Hamiltonian map as in section
    \ref{sCD}, the corresponding quantum propagator is defined
    \[\calU_N(\phi_H)=e^{\frac{i}{\hbar}\Op_N(H)}=e^{2\pi i N\Op_N(H)}\]
    This is a unitary operator (because $\Op_N(H)$ is Hermitian), and it satisfies the Erogov identity in the
    semi-classical limit  \cite{BD3}. That is, for any smooth observable $f\in
    C^\infty(\bbT^{2d})$
    \[\norm{\calU_N(\phi_H)^*\Op_N(f)\calU_N(\phi_H)-\Op_N(f\circ
    \phi_H)}=O_{f}(\frac{1}{N^2})\]
    (see appendix \ref{sEGOROV} for more details).

    For $A\in \Sp(2d,\bbZ)$ which satisfies
    certain parity conditions, one can assign
    unitary operators $\calU_N(A)$, acting on $\calH_N$
    satisfying ``Exact Egorov"  \cite{BD1,K}. That is,
    for all observables $f\in
    C^\infty(\mathrm{\bbT}^{2d})$
    \[\calU_N(A)^*\Op_N(f)\calU_N(A)=\Op_N(f\circ A).\]

    The quantization of the perturbed map $\phi=A\circ \phi_H$ is the
    composition of the quantum propagators for $A$ and $\phi_H$ respectively,
      \[\calU_N(\Phi)=\calU_N(A)\calU_N(\phi_{H}).\]
    Consequently, this is again a unitary operator satisfying the
    Egorov identity in the semi-classical limit
    \[\norm{\calU_N(\Phi)^*\Op_N(f)\calU_N(\Phi)-\Op_N(f\circ
    \phi)}=O_{f}(\frac{1}{N^2})\]

\subsection{Limiting measures}
    One way to study the eigenstates of the propagator is through
    their corresponding Wigner distributions.
    The Wigner distribution $\calW_N(\psi)$ (of a quantum states $\psi\in\calH_N$),
    is a distributions on the phase space $\bbT^{2d}$, that
    assigns to a smooth observable its expectation value
    \[\calW_N(\psi)(f)=\langle\Op_N(f)\psi,\psi\rangle.\]

    We call a measure $\mu$ on $\bbT^{2d}$ a limiting quantum
    measure, if there exists a sequence $\psi_j\in\calH_{N_j}$ of eigenstates of $\calU_{N_j}(\Phi)$,
    such that the corresponding Wigner distributions $\calW_{N_j}(\psi_j)\stackrel{w^*}\to \mu$
    converge to this measure when $N_j\to \infty$.

\section{Dynamics on invariant manifolds}\label{sCLAS}
\subsection{Invariant manifolds}\label{sINV}
    Let $A\in\Sp(2d,\bbZ)$ be a symplectic matrix with integer
    coefficients, and assume that $A$ has no eigenvalues
    of modulus $1$. The natural left action of $A$ on $\bbR^{2d}$ ($x\mapsto Ax$) preserves the lattice $\bbZ^{2d}$,
    and induce discrete time dynamics on the torus
    $\bbT^{2d}=\bbR^{2d}/\bbZ^{2d}$ which is invertible area preserving
    and stably ergodic and mixing.

    There is also a natural right action of $A$ on the rational vector space $\bbQ^{2d}$ ($n\mapsto nA$).
    For any rational subspace $V\subset\bbQ^{2d}$ that is preserved by $A$
    there is a corresponding  closed connected subgroup $X_0\subseteq \bbT^{2d}$
    invariant under the induced dynamics. Let $\Lambda=V\cap
    \bbZ^{2d}$ be the lattice obtained by taking the integral
    points of $V$, then the corresponding subgroup is
    \[X_0=\set{x\in\bbT^{2d}|e_n(x)=1,\;\forall n\in \Lambda}\]
    Let $\dim V=d_0$, then $X\cong \bbT^{2d-d_0}$ is a subtorus of codimension $d_0$.
    Furthermore, for any fixed point of the dynamics
    $\xi\in\bbT^{2d}$  (i.e., $A\xi\equiv\xi\pmod{1}$),
    the manifold
    \[X_{\xi}=\xi+X_0=\set{x\in\bbT^{2d}|e_n(x)=e_n(\xi),\;\forall n\in
    \Lambda}\]
    is also a connected closed sub-manifold that is preserved by the induced dynamics.
    We say that these manifolds are co-isotropic,
    when the invariant subspace $V$ is isotropic with respect to the symplectic
    form.

    \subsection{Stable ergodicity}
    It is a well known result of Anosov \cite{An}, that hyperbolic automorphisms of the torus
    are stably ergodic. We now show that the
    restriction of $A$ to $X_\xi$ can be identified with a hyperbolic automorphisms
    of $\bbT^{2d-d_0}$ implying the following:
   \begin{prop}\label{pSTAB}
    The restriction of $A$ to $X_\xi$ preserves the volume measure $dm_{X_\xi}$
    concentrated on $X_\xi$, and is stably ergodic
    (with respect to $dm_{X_\xi}$).
   \end{prop}
    \begin{proof}
    Since the map on $X_\xi$ is just shifting by $\xi$ of the
    map on $X_0$, it is sufficient to show this for $X_0$.
    Let $W=\set{x\in \bbR^{2d}|n\cdot x=0,\;\forall n\in
    V}$, then $W$ is invariant under the left action of $A$.
    Let $\Omega=W\cap\bbZ^{2d}$ be the lattice of integral
    points, then $\Omega$ is of rank
    $2d-d_0$ (because $V$ is rational). The natural injection $W\hookrightarrow\bbR^{2d}$ induces an
    imbedding $W/\Omega\hookrightarrow \bbT^{2d}$ with image $X_0$.
    Therefore, $X_0\cong W/\Omega$ and it is sufficient to
    prove the proposition for the action of $A$ on $W/\Omega$.

    Fix an integral basis $n_1,\ldots,n_{2d-d_0}$ for $\Omega$,
    and let $B\in\GL(2d-d_0,\bbZ)$ be the matrix corresponding to the
    action of $A$ on $\Omega$.
    By taking coordinates in the integral basis we can identify
    $W/\Omega\cong
    \bbR^{2d-d_0}/\bbZ^{2d-d_0}=\bbT^{2d-d_0}$, and under this
    identification the action on $\bbT^{2d-d_0}$ is the automorphism induced by the natural
    action of $B$ on $\bbR^{2d-d_0}$. This map is hyperbolic (since eigenvalues of $B$ are also eigenvalues of $A$)
    and area preserving (since $\det(B)=\pm 1$) and hence stably ergodic.
    \end{proof}

\subsection{Perturbation preserving invariant
manifolds}\label{sPER}
    We now consider the perturbed map $\Phi=A\circ \phi_H$.
    We require that the manifolds $X_\xi$ remain invariant under
    by the perturbed map. For that reason, we choose our
    Hamiltonian $H$ such that the functions
    $e_n(x),\;n\in\Lambda$ defining the manifolds $X_{\xi}$ are
    constants of motion for the Hamiltonian flow.
    This is equivalent to requiring that the Poisson brackets $\{H,e_n\}=0$
    for all $e_n,\;n\in\Lambda$. In terms of the Fourier coefficients
    this is equivalent to the requirement that $\hat{H}(m)=0$ unless
    $m\in\Lambda^\bot=\set{m\in\bbZ^{2d}|\omega(n,m)=0,\; \forall n\in\Lambda}$.

    By proposition \ref{pSTAB} (replacing the Hamiltonian $H$ by $\epsilon H$ if
    necessary) we can insure that the perturbed map and its restriction to the
    $X_\xi$'s remains ergodic.

    %\begin{prop}
%    Let $\Lambda^\bot=\set{m\in\bbZ^{2d}|\omega(n,m)=0,\;\forall
%    n\in\Lambda}$. Then, $\{g,e_n\}=0$ for any $n\in\Lambda$
%    if and only if $\hat{g}(n)=0$ for $n\not\in\Lambda^\bot$.
%    \end{prop}
%    \begin{proof}
%    Decompose $g$ in to Fourier series $g=\sum \hat{g}(m)e_m$.  Then
%    \[\{g,e_n\}=\sum_m \hat{g}(m) \omega(n,m)e_{n+m}\]
%    This is again a smooth function on $\bbT^{2d}$ and hence
%    vanishes if and only if all its Fourier coefficients $\hat{g}(m)\omega(n,m)=0$.
%    \end{proof}

\section{Scarring on invariant manifolds}\label{sSCAR}
    Let $A\in\Sp(2d,\bbZ)$ be as in the previous section,
    let $V\subseteq \bbQ^{2d}$ be an invariant isotropic rational
    subspace of dimension $1\leq d_0\leq d$ and let $\Lambda=V\cap \bbZ^{2d}$.
    For any fixed point $\xi\in\bbT^{2d}$ let
    $X_\xi=\set{x\in\bbT^{2d}|e_n(x)=e_n(\xi),\;\forall n\in\Lambda}$ be the corresponding (co-isotropic)
    invariant manifold as in section \ref{sINV}. Let $\Phi=A\circ
    \phi_H$ be the perturbed map leaving the manifolds $X_\xi$
    invariant as in section \ref{sPER}, and let $\calU_N(\Phi)$ be
    its quantization.

\subsection{Proof of Theorem \ref{tSCARR}}
    In order to prove theorem \ref{tSCARR}, we first show
    that the spaces $\calH_{N,\xi}=\set{\psi|\Op_N(e_n)\psi=e_n(\xi)\psi,\forall
    n\in \Lambda}$ are of dimension $N^{d-d_0}$ and are preserved
    by $\calU_N(\phi)$. Then we show the localization of states
    from $\calH_{N,\xi}$.

    Consider the family of operators $\calA_N=\set{\Op_N(e_n)
    |n\in\Lambda}$.
    This is a commutative family of unitary operators (recall $\Lambda$ is
    isotropic).
    We can thus decompose
    the Hilbert space $\calH_N$ into joint eigenspaces
    \[\calH_N=\bigoplus_{\widehat{\Lambda/N\Lambda}}\calH_\lambda\]
    where the sum is over characters of $\Lambda/N\Lambda\cong \calA_N$.

    \begin{lem}
    Any character $\lambda$ of $\Lambda$ of order $N$, is of the form
    $\lambda(n)=e(\frac{\omega(n,m)}{N})$ for some $m\in\bbZ^{2d}$.
    \end{lem}
    \begin{proof}
    Notice, that if we have an integral basis $n_i$ $i=1,\ldots
    n_{d_0}$ for $\Lambda$ and $m_i\in\bbZ^{2d}$, such that
    $\omega(n_i,m_j)=\delta_{i,j}$ then we are done. Indeed, since we
    assume the character is of order $N$, for the basis elements
    $\lambda(n_i)=e(\frac{k_i}{N})$ for some $k_i\in\bbZ$. Then for $m=\sum_i
    k_i m_i\in\bbZ^{2d}$
    \[\lambda(n)=\lambda(\sum_i a_i n_i)=e(\frac{\sum_{i,j} \omega(a_i n_i,k_jm_j)}{N})=e(\frac{\omega(n,m)}{N}).\]

    We now proceed by induction on $\dim V=\rank (\Lambda)$ and construct such an integral basis.
    If $\dim
    V=1$ then choose any primitive vector $n\in\Lambda$ (i.e.,
    $\gcd(n)=1$). Then there is $m\in\bbZ^{2d}$ such that
    $\omega(n,m)=\gcd(n)=1$.

    Now for $\dim(V)=l>1$ assume that
    $n_1,\ldots n_{l-1}$ are an integral basis with $m_1,\ldots
    m_{l-1}\in\bbZ^{2d}$ such that $\omega(n_i,m_j)=\delta_{i,j}$.
    Take an element $\tilde n_l\in\Lambda$ linearly independent on
    $n_1,\ldots n_{l-1}$, and let
    \[n_l=\tilde n_l-\sum_{i=1}^{l-1} \omega(\tilde n_l,m_i)n_i.\]
    Then $n_1,\ldots n_l$ are still linearly independent and $\omega(n_l,m_i)=0$ for $i<l$.
    We can assume
    that $\gcd{n_l}=1$ (otherwise divide by any common factors), so there is
    $\tilde m_l\in\bbZ^{2d}$ such that $\omega(n_l,\tilde m_l)=1$. Take
    \[m_l=\tilde m_l-\sum_{i=1}^{l-1} \omega(\tilde n_i,m_l)m_i\]
    so indeed $\omega(m_i,n_j)=\delta_{i,j}$ for any $1\leq i,j\leq l$.

    %It now remains to show that these $n_i$'s are indeed an integral basis. For any
%    $n\in\Lambda$ there are $a_i\in\bbQ$ such that $n=\sum_i a_in_i$
%    (because the $n_i$'s are linearly independent and hence span $V$).
%    The coefficients in this decomposition satisfy $a_i=\omega(n,m_i)\in\bbZ$, and hence
%    integral.
    \end{proof}

    \begin{cor}
    The dimension of all eigenspaces satisfy
    \[\dim \calH_\lambda=N^{d-d_0}\]
    \end{cor}
    \begin{proof}
    Any character of $\Lambda/N\Lambda$ is of the form
    $\lambda_m(n)=e(\frac{\omega(m,n)}{N})$ for some $m\in\bbZ^{2d}$.
    Hence, the operator $\Op_N(e_m)$ sends the space $\calH_{\lambda_0}$ into
    the space $\calH_{\lambda_m}$ with inverse map $\Op_N(e_{-m})$.
    Hence for all characters $\dim
    \calH_\lambda=\dim\calH_{\lambda_0}$ and since there are $N^{d_0}$
    characters then
    \[N^{d}=\dim\calH_N=\sum_\lambda \dim\calH_\lambda=N^{d_0}\dim\calH_{\lambda_0} \]
    concluding the proof.
    \end{proof}

    Let $\xi\in\bbT^{2d}$ be a fixed point for $A$ and
    consider the character of $\Lambda$ given by
    $\lambda_{\xi}(n)=e_n(\xi)$. For $R=\det(A-I)$ we know that $R\xi\in\bbZ^{2d}$.
    Therefore, if $N$ is divisible by $R$ then the character $\lambda_{\xi}$ is a
    character of $\Lambda/N\lambda$ and the space
    $\calH_{N,\xi}=\calH_{\lambda_\xi}$, we thus showed
    \begin{cor}
    $\dim \calH_{N,\xi}=N^{d-d_0}.$
    \end{cor}
    \begin{prop}
    The spaces $\calH_{N,\xi}$ are preserved by $\calU_N(\Phi)$.
    \end{prop}
    \begin{proof}
    Recall, that the quantum propagator
    $\calU_N(\Phi)=\calU_N(A)\calU_N(\phi_{H})$ is the composition of
    $\calU_N(A)$ and the quantization of the hamiltonian flow $\calU_N(\phi_{H})=e^{2\pi i N\Op_N(H)}$.
    We will show that both $\calU_N(\phi_{H})$ and $\calU_N(A)$
    preserve $\calH_{N,\xi}$.

    The quantum observable $\Op_N(H)=\sum_{m\in\Lambda^\bot}\hat{H}(m)\Op_N(e_m))$
    and hence commutes with $\Op_N(e_n)$ for all
    $n\in\Lambda$. Consequently, $\calU_N(\phi_{H})$ also commutes with $\Op_N(e_n)$
    for all $n\in\Lambda$, and preserves the eigenspace space $\calH_{N,\xi}$.
     %=\set{\psi\in\calH_N|\Op_N(e_n)\psi=e_n(\xi)\psi,\;\forall n\in\Lambda}$.

    Next, for any $n\in\Lambda$ and any $\psi\in\calH_{N,\xi}$ we have
    \[\Op_N(e_n)\calU_N(A)\psi=\calU_N(A)\Op_N(e_{nA})\psi\]
    Because $\Lambda$ is invariant, $nA\in\Lambda$ as
    well so $\Op_N(e_{nA})\psi=e_{nA}(\xi)\psi$. Because $\xi$ is a
    fixed point of $A$ then $e_{nA}(\xi)=e_n(A\xi)=e_n(\xi)$ so that
    \[\Op_N(e_n)\calU_N(A)\psi=e_n(\xi)\calU_N(A)\psi,\]
    and the space $\calH_{N,\xi}$ is also preserved by $\calU_N(A)$.
\end{proof}

    Let $f\in C^\infty(\bbT^{2d})$ and consider the restriction of
    $f$ to the manifold $X_\xi$. For any $x\in X_\xi$ and any $m\in\Lambda$ we have
    $e_m(x)=e_m(\xi)$, we can thus write
    \[f(x)=\sum_m \hat{f}(m)e_m(x)=\sum_{n\in\Sigma}f^\sharp(n)e_n(x),\]
    where $f^\sharp(n)=\sum_{m\in\Lambda}\hat{f}(n+m)e_m(\xi)=\int_X f(x) e_{-n}(x)
    dm_{X_{\xi}}$, and $\Sigma\subseteq \bbZ^{2d}$ is any set of
    representatives for $\bbZ^{2d}/\Lambda$.

    \begin{lem}\label{lREP}
    There is a choice of representatives $\Sigma\subseteq \bbZ^{2d}$
    for $\bbZ^{2d}/\Lambda$ such that for any $n\in\Sigma$ and
    $m\in\Lambda$ we have $|\omega(n,m)|\ll \norm{n+m}^2$
    \end{lem}
    \begin{proof}
    Let $U\subseteq \bbQ^{2d}$ be the orthogonal complement of $V$ (for the standard inner product).
    Then there is an integer $D\in\bbZ$ (depending on $V$) such
    that the images of the projection maps $P_V(D\bbZ^{2d})$ and
    $P_U(D\bbZ^{2d})$ lie inside $\bbZ^{2d}$.

    Denote by  $[D]^{2d}=\{1,\ldots,D\}^{2d}$, and think of it as a set of representatives for $\bbZ^{2d}/D\bbZ^{2d}$.
    Then our set of representatives for $\bbZ^{2d}/\Lambda$ is taken to be
    \[\Sigma= \set{P_U(m-r)+r|m\in \bbZ^{2d}, r\in[D]^{2d}\mbox{ with } r\equiv m\pmod D}\]
    This is indeed a set of representatives, since any
    $k\in\bbZ^{2d}$ can be written uniquely as
    $k=P_V(k-r)+P_U(k-r)+r$ with $r\in [D]^{2d},\;r\equiv k\pmod{D}$ and $P_V(k-r)\in\Lambda$.

    Now for any $n\in\Sigma$ and $m\in\Lambda$ let $r\equiv n+m\pmod{D}$ in $[D]^{2d}$.
    Then $m=P_V(n+m-r)$ and $n=P_U(n+m-r)+r$, hence
    \[|\omega(n,m)|\leq \norm{n}\norm{m}\ll
    \norm{n+m-r}^2\ll\norm{n+m}^2.\]
\end{proof}
    The following proposition concludes the proof of theorem
    \ref{tSCARR}.
\begin{prop}
    Let $\psi_{j}\in \calH_{N_j,\xi}$ be a sequence of states such that
    the Wigner distributions
    $\calW_{N_j}(\psi_j)\stackrel{w^*}{\to} \mu$ converge
    weak$^*$ to some measure $\mu$ on $\bbT^{2d}$,
    then $\mu$ is supported on $X_{\xi}$.
\end{prop}

\begin{proof}
    Let $\mu$ be such a limiting measure. In order to show $\supp
    \mu\subseteq X_{\xi}$ it is sufficient to show that any smooth
    function $f$
    vanishing on $X_{\xi}$, satisfies $\mu(f)=\int f d\mu=0$.

    Fix a smooth function $f$ vanishing on $X_{\xi}$. Let $\Sigma$ be a set of representatives for $\bbZ^{2d}/\Lambda$
    as in lemma \ref{lREP}. Then, for any $n\in\Sigma$
    \[f^\sharp(n)=\sum_{m\in\Lambda}\hat{f}(n+m)e_m(\xi)=\int_{X_\xi} f(x) e_{-n}(x) dm_{X_{\xi}}=0.\]

      For any fixed $N=N_j$, and $\psi=\psi_j\in \calH_{N,\xi}$
    \[\langle
    \Op_N(f)\psi,\psi\rangle=\sum_{n\in \Sigma}\sum_{m\in\Lambda}\hat{f}(n+m)\langle
    \Op_N(e_{n+m})\psi,\psi\rangle\]
    Replace
    \[\langle
    \Op_N(e_{n+m})\psi,\psi\rangle=e_{2N}(\omega(n,m))e_m(\xi)\langle
    \Op_N(e_n)\psi,\psi\rangle\]
    to get
    \[\langle
    \Op_N(f)\psi,\psi\rangle=\sum_{n\in \Sigma}\langle
    \Op_N(e_n)\psi,\psi\rangle\sum_{m\in\Lambda}\hat{f}(n+m)e_m(\xi)
    e_{2N}(\omega(n,m)).\]
    Since we know $f^\sharp(n)=\sum_{m\in\Lambda}\hat{f}(n+m)e_m(\xi)=0$ we can subtract it to get
    \[\langle
    \Op_N(f)\psi,\psi\rangle=\!\!\!\sum_{n\in \Sigma, m\in\Lambda}\langle
    \Op_N(e_n)\psi,\psi\rangle\hat{f}(n+m)e_m(\xi)(
    e_{2N}(\omega(n,m))-1)\]
    We can thus bound
    \[\langle
    \Op_N(f)\psi,\psi\rangle\leq \sum_{n\in\Sigma}\sum_{m\in\Lambda}|\hat{f}(n+m)||(
    e_{2N}(\omega(n,m))-1)|\]
    and since
    $|e_{2N}(\omega(n,m))-1)|\ll\frac{|\omega(n,m)|}{N}\ll \frac{\norm{n+m}^2}{N}$
    we get
    \[|\langle
    \Op_N(f)\psi,\psi\rangle|\ll \frac{1}{N}\sum_{m\in
    \bbZ^{2d}}|\hat{f}(m)|\norm{m}^2\ll_f \frac{1}{N}\]
    and indeed
    \[\mu(f)=\lim_{N_j\to \infty}\langle
    \Op_{N_j}(f)\psi_j,\psi_j\rangle=0\]
\end{proof}

\subsection{Proof of theorem \ref{tERG}} The proof of theorem \ref{tERG} follows the lines
of the proof of the Quantum Ergodicity Theorem ~\cite{BD,Z}. The
first ingredient is showing that on average the sates in
$\calH_{\xi}$ are evenly distributed in $X_{\xi}$ (proposition
\ref{pAVR}). Then we use the ergodicity of the restricted map on
$X_{\xi}$ to bound the variance (theorem \ref{tVAR}). Theorem
\ref{tERG} is then derived by a standard diagonalization argument.

\begin{prop}\label{pAVR}
For any integer $N$ divisible by $R$, let
$\{\psi_j\}_{j=1}^{N^{d-d_0}}$ be an orthonormal basis for
$\calH_{N,\xi}\subset \calH_N$. For any smooth $f\in
C^\infty(\bbT^{2d})$ as $N\to \infty$ the average
\[\frac{1}{\dim \calH_{N,\xi}}\sum_j \langle
\Op_N(f)\psi_j,\psi_j\rangle\longrightarrow \int_{X_{\xi}}f
dm_{X_\xi}.\]
\end{prop}

\begin{proof}
    It is sufficient to show this holds for $f=e_n$ for all
    $n\in\bbZ^{2d}$. Note that
    \[\int_{X_\xi}e_n dm_{X_{\xi}}=\left\lbrace\begin{array}{cc} e_n(\xi)
    & n\in \Lambda\\ 0 & \mbox{otherwise}\end{array}\right..\]

    Now for $n\in\Lambda$ by definition
    $\Op_N(e_n)\psi_j=e_n(\xi)\psi_j$ so this is trivial. For
    $n\not\in \Lambda$ there are two possible cases, either there is
    $m\in \Lambda$ such that $\omega(m,n)\neq 0$ or there is not.

    In the case that such $m\in\Lambda$ exists, since
    $\Op_N(e_m)\psi_j=e_m(\xi)\psi_j$ we can write
    \begin{eqnarray*}
   \sum_j \langle
    \Op_N(e_n)\psi_j,\psi_j\rangle &=& \sum_j \langle
    \Op_N(e_n)\Op_N(e_m)\psi_j,\Op_N(e_m)\psi_j\rangle\\
    &=& \sum_j \langle
    \Op_N(e_{-m})\Op_N(e_n)\Op_N(e_m)\psi_j,\psi_j\rangle\\
    &=& e\big(\frac{\omega(n,m)}{N}\big)\sum_j \langle
    \Op_N(e_n)\psi_j,\psi_j\rangle .
    \end{eqnarray*}
    We can assume $N$ large enough so
    that $e(\frac{\omega(n,m)}{N})\neq 1$ implying that $\frac{1}{\dim
    \calH_{N,\xi}}\sum_j \langle \Op_N(e_n)\psi_j,\psi_j\rangle=0$.

    If on the other hand $\omega(m,n)=0$ for all $m\in\Lambda$, then
    the operator $\Op_N(e_n)$ commutes with all the operators
    $\Op_N(e_m),\;m\in\Lambda$ and hence preserves the space
    $\calH_{N,\xi}$. We can thus look at the restriction of
    $\Op_N(e_n)$ to $\calH_{N,\xi}$ and the sum $\sum_j \langle
    \Op_N(e_n)\psi_j,\psi_j\rangle=\Tr(\Op_N(e_n)|_{\calH_{N,\xi}})$
    is the trace of this restricted operator. Now, fix $k\in\bbZ^{2d}$
    such that $\omega(n,k)\neq 0$ but $\omega(m,k)=0$ for any
    $m\in\Lambda$. Such a vector exists since we can take $\tilde{k}$
    to be any vector with $\omega(n,\tilde k)\neq 0$ and then define
    $k=\tilde{k}-\sum \omega(\tilde{k},m_i)m_i$ for some integral
    basis $m_i$ of $\Lambda$. The operator $\Op_N(e_k)$ also preserves
    the space $\calH_{N,\xi}$ and so
     \begin{eqnarray*}
     \Tr(\Op_N(e_n)|_{\calH_{N,\xi}})&=&\Tr(\Op_N(e_{-k})\Op_N(e_n)\Op_N(e_k)|_{\calH_{N,\xi}})\\
    &=&e(\frac{\omega(n,k)}{N})\Tr(\Op_N(e_n)|_{\calH_{N,\xi}})
     \end{eqnarray*}
    Again we can assume that $N$ is sufficiently large so that
    $e(\frac{\omega(n,k)}{N})\neq 1$, implying that
    $\Tr(\Op_N(e_n)|_{\calH_{N,\xi}})=0$.
\end{proof}

    For any $N$ (divisible by $R$) we fix an orthonormal basis of eigenstates
    $\psi_i\in\calH_{N,\xi},\;i=1,\ldots, N^{d-d_0}$.
    For any smooth function $f\in C^\infty(\bbT^{2d})$ we define the quantum variance (in
    this basis) as
    \[\sigma^2_N(f)=\frac{1}{\dim\calH_{N,\xi}}\sum_i|\langle
    \Op_N(f)\psi_i,\psi_i\rangle-\int_{X_\xi} fdm_{X_{\xi}}|^2\]

    \begin{thm}\label{tVAR}
    For $f\in C^\infty(\bbT^{2d})$, $\lim_{N\to
    \infty}\sigma_N^2(f)=0$.
    \end{thm}
\begin{proof}
     With out loss of generality we can assume that $\int_X
    fdm_{X_{\xi}}=0$. For $T>0$ denote by
    $f^T=\frac{1}{T+1}\sum_{t=0}^T f\circ \phi^t$ the time average of $f$.
    For fixed $t$ we have
    \[\norm {\calU_N(\Phi)^{-t}\Op_N(f)\calU_N(\Phi)^t-\Op_N(f\circ
    \phi^t)}=O_{f,t}(\frac{1}{N^2})\]
    Hence for any eigenstate $\psi$ of $\calU_N(\Phi)$ we have
    \begin{equation}\label{eEG}
    |\langle \Op_N(f^T)\psi,\psi\rangle-\langle
    \Op_N(f)\psi,\psi\rangle|=O_{f,T}(\frac{1}{N^2}).
    \end{equation}

    Using Cauchy-Schwarz inequality
    \[|\langle
    \Op_N(f^T)\psi,\psi\rangle|^2\leq
    \norm{\Op_N(f^T)\psi}^2\norm{\psi}^2=\langle
    \Op_N(f^T)^*\Op_N(f^T)\psi,\psi\rangle,\]
    we get
    \begin{equation}\label{eCS}
    \sum_j |\langle \Op_N(f^T)\psi_j,\psi_j\rangle|^2\leq \sum_j
    \langle
    \Op_N(f^T)^*\Op_N(f^T)\psi_j,\psi_j\rangle
    \end{equation}
    Now, the estimate (\ref{ePROD}) for $\Op_N(f^T)^*\Op_N(f^T)$ and $\Op_N(|f^T|^2)$
    implies
    \begin{equation}\label{ePR}
    \frac{1}{\dim\calH_{N,\xi}}\sum_j \langle
    \Op_N(f^T)^*\Op_N(f^T)\psi_j,\psi_j\rangle=
    \end{equation}
    \[ =\frac{1}{\dim\calH_{N,\xi}}\sum_j \langle
    \Op_N(|f^T|^2)\psi_j,\psi_j\rangle+O_{T,f}(\frac{1}{N}).\]

    From the estimates given in (\ref{eEG}),(\ref{eCS}) and (\ref{ePR}) we get
    \[\sigma_N^2(f)\leq \frac{1}{\dim\calH_{N,\xi}}\sum_j \langle
    \Op_N(|f^T|^2)\psi_j,\psi_j\rangle+O_{T,f}(\frac{1}{N}),\]
    and in the limit $N\to\infty$
    \begin{equation}\label{eVAR}
    \limsup_{N\to \infty}\sigma_N^2(f) \leq \int_{X_{\xi}}|f^T|^2 d m_{X_{\xi}}=\norm{f^T}^2_{L^2(X_{\xi})}
    \end{equation}
    This is true for any $T>0$. However, since the map
    $\Phi=A\circ \phi_{H}$ induces ergodic dynamics on the
    manyfold $X_{\xi}$, then in the limit $T\to \infty$ we have
    \[f^T\stackrel{L^2(X_{\xi})}{\longrightarrow} 0,\]
    and since the left hand side of (\ref{eVAR}) has no dependence on
    $T$ indeed $\lim_{N\to \infty}\sigma_N^2(f)=0$
\end{proof}
    We now give the proof of theorem \ref{tERG} from theorem
    \ref{tVAR}.
\begin{proof}
    For each $n\in \bbZ^{2d}$, and $N\in\bbN$ denote by
    $\sigma^2_N(n)=\sigma_N^2(e_n)$. Then theorem \ref{tVAR} implies that
    $\sigma_N(n)\stackrel{N\to\infty}{\To}0$. Now for fixed $n\not\in\Lambda$ use Chebyshev's inequality
    to get that
    \[\frac{1}{N^{d_0}}\sharp\set{i\big||\calW_N(\psi_i)(e_n)|\geq
    \sqrt{\sigma_N(n)}}\leq \sigma_N(n).\]

    For any integer $M>0$ define the set
    \[J_N(M)=\set{i|\exists n\not\in \Lambda,\;\norm{n}\leq M,\;|\calW_N(\psi_i)(e_n)|\geq \sqrt{\sigma_N(n)}}.\]
    For fixed $M$, we have that $\frac{\sharp J_N(M)}{N^{d_0}}\leq\sum_{\norm{n}\leq M}\sigma_N(n)$ and in the
    limit $\lim_{N\to\infty}\frac{\sharp{J_N(M)}}{N^{d_0}}=0$.
    Consequently, there is a sequence $M_N\to\infty$ such that
    $\lim_{N\to\infty}\frac{\sharp{J_N(M_N)}}{N^{d_0}}=0$, and the
    sets
    \[S_N=\{1,\ldots,N^{d_0}\}-J_N(M_N),\]
    are of density one in $\{1,\ldots,N^{d_0}\}$.

    Now let $f\in C^\infty(\bbT^{2d})$ be a smooth function. Fix $\epsilon>0$ and let $M$ be sufficiently large so that
    $\sum_{\norm{n}>M}|\hat{f}(n)|\leq \epsilon$. Fix $N_0$ sufficiently large so
    that $M_{N}>M$ for $N>N_0$.
    Then, for any $\psi_i$ we have that
    \[|\calW_N(\psi_i)(f)-\int_{X_\xi}fdm_{X_\xi}|\leq
    \sum_{n\not\in\Lambda,\norm{n}<M}\calW_N(\psi_i)(e_n)+\epsilon.\]
    But for $i\in S_N$ and $\norm{n}\leq M\leq M_N$ by definition
    $\calW_N(\psi_i)(e_n)\leq \sqrt{\sigma_N(n)}$. We thus get that
    \[|\calW_N(\psi_i)(f)-\int_{X_\xi}fdm_{X_\xi}|\leq
    \sum_{0\neq\norm{n}<M}|\hat{f}(n)|\sqrt{\sigma_N(n)}+\epsilon.\]
    Taking $N\to\infty$ we get
    \[\limsup_{N\to\infty}|\calW_N(\psi_i)(f)-\int_{X_\xi}fdm_{X_\xi}|\leq \epsilon,\]
    implying that indeed
    $\calW_N(\psi_i)(f)\stackrel{N\to\infty}{\longrightarrow}\int_{X_\xi}fdm_{X_\xi}$.

    \end{proof}
\appendix

\section{Proof of Egorov}\label{sEGOROV}
    The (semi classical) Egorov theorem, is a well known result for quantization of
    Hamiltonian flows on $\bbR^{2d}$. For Hamiltonian flows on
    $\bbT^{2d}$ the proof is analogous and is described in \cite{BD3}. For the sake of
    completeness, we give a short proof along the same lines.

    Recall, that given a real valued smooth function $g\in\bbT^{2d}$,
    the associated Hamiltonian flow, $\phi_g^t:\bbT^{2d}\to\bbT^{2d}$
    satisfies the differential equations:
    \begin{equation}\label{eHAM}
    \frac{d}{dt}(f\circ\phi)=\{g,f\circ\phi^t\},\quad \forall f\in
    C^\infty(\bbT^{2d}),
    \end{equation}
    and note that the quantum propagator $\calU_N(\phi_g^t)=e^{2\pi iN{\Op_N(g)}t}$  corresponding to this flow
    satisfies
    \begin{equation}\label{ePROP}
    \frac{d\calU_N(\phi_g^t)}{dt}=2\pi
    iN\Op_N(g)\calU_N(\phi_g^t)=2\pi
    iN\calU_N(\phi_g^t)\Op_N(g).
    \end{equation}

    The main ingredient in the proof, is the connection between the
    Poisson brackets and quantum commutators.
    \begin{lem}\label{lPOISCOM}
    For any $f\in
    C^\infty(\bbT^{2d})$ let
    $c_f=\norm{(-\triangle)^{d+2} f}_\infty$.
    Then, there is a constant $C$, such that for any $f,g\in C^\infty(\bbT^{2d})$,
        \[\norm{\Op_N(\{g,f\})-[2\pi iN\Op_N(g),\Op_N(f)]}\leq C\frac{c_gc_f}{N^2}, \]
    \end{lem}
    \begin{proof}
        For any $n,m\in\bbZ^{2d}$,
        \[\Op_N(\{e_m,e_n\})-[2\pi
        iN\Op_N(e_n),\Op_N(e_m)]=\]
        \[=[(2\pi)^2\omega(m,n)-
        2\pi 2N\sin(\frac{2\pi\omega(m,n)}{2N})]\Op_N(e_{n+m}).\]
        Hence,
        \[\norm{\Op_N(\{e_m,e_n\})-[2\pi iN\Op_N(e_n),\Op_N(e_m)]}
        \leq\frac{4\pi^3\norm{m}^3\norm{n}^3}{N^2}.\]
        Decomposing $g$ and $f$ into Fourier series, noting that
        $|\hat{f}(n)|$ and $|\hat{g}(n)|$ are bounded by
        $\frac{c_f}{(2\pi\norm{n})^{2d+4}}$ and
        $\frac{c_g}{(2\pi\norm{n})^{2d+4}}$, one gets
        \[\Op_N(\{g,f\})-[2\pi
        iN\Op_N(g),\Op_N(f)]\leq C \frac{c_gc_f}{N^2},\]
        with $C=|\sum_n\frac{1}{(2\pi\norm{n})^{2d+1}}|^2$.
    \end{proof}
    For $f\in C^\infty(\bbT^{2d})$ smooth, its composition $f\circ\phi_g^s$ with
    the Hamiltonian flow is
    also smooth. We can thus consider
    \[c_{f\circ\phi_g^s}=\norm{(-\triangle)^{d+2} (f\circ\phi_g^s)}_\infty,\]
    and let
    \[C_{f,g}(t)=C\cdot c_g\cdot \sup_{0\leq s\leq t}( c_{f\circ \phi_g^s}).\]
    \begin{thm}[Egorov] For all $f,g\in
    C^{\infty}(\bbT^{2d})$ we have
        \[\norm{\calU_N(\phi_g^t)^*\Op_N(f)\calU_N(\phi_g^t)-\Op_N(f\circ
        \phi_g^t)}\leq \frac{tC_{f,g}(t)}{N^2}.\]
    \end{thm}
    \begin{proof} Denote by $B(s)=\calU_N(\phi_g^s)\Op_N(f\circ
        \phi_g^s)\calU_N(\phi_g^s)^*$.
        Since conjugating by a unitary matrix doesn't change the
        norm, it is equivalent to bound the norm of
        \[\norm{\calU_N(\phi_g^t)\Op_N(f\circ
        \phi_g)\calU_N(\phi_g^t)^*-\Op_N(f)}=\norm{B(t)-B(0)}.\]
        %\[=e^{-2\pi iN{\Op_N(g)}t}\Op_N(f\circ
        %\phi_g^t)e^{2\pi iN{\Op_N(g)}t}.\]

        Differentiate $B(s)$, recalling (\ref{ePROP}) and (\ref{eHAM})
        %that $\frac{d\calU_N(\phi_g^t)}{dt}=2\pi i\Op_N(g)\calU_N(\phi_g^t)$ and that
        %$\frac{d\Op_N(f\circ\phi_g^t)}{dt}=\Op_N(\{g,f\circ\phi_g^t\})$,
        to get
        \[B'(s)=-[2\pi
        iN\Op_N(g),B(s)]+\calU_N(\phi_g^s)^*\frac{d(\Op_N(f\circ\phi_g^s))}{ds}\calU_N(\phi_g^s)^*\]
        \[=\calU_N(\phi_g^s)^*(\Op_N(\{g,f\circ\phi_g^s\})-[2\pi
        iN\Op_N(g),\Op_N(f\circ\phi_g^s)])\calU_N(\phi_g^s).\]
        Consequently, for any $0\leq s\leq
        t$ we can bound (using lemma \ref{lPOISCOM})
        $\norm{B'(s)}\leq \frac{C_{f,g}(t)}{N^2}$, implying that
        \[\norm{B(t)-B(0)}=\norm{\int_0^tB'(s)ds}\leq
       \frac{tC_{f,g}(t)}{N^2},\]
       as claimed.
    \end{proof}

% ----------------------------------------------------------------
%%GATHER{varbib.bib}   % For Gather Purpose Only
%%GATHER{Paper.bbl}  % For Gather Purpose Only
%\bibliographystyle{amsplain}
%\bibliography{varbib}
\end{document}